\documentclass[a4paper,11pt]{article}
\pdfoutput=1

\usepackage{amsmath}
\usepackage{jheppub}
\usepackage{graphicx}
\usepackage{amsbsy,amssymb}
\usepackage{mathtools}
\usepackage{xspace}
\usepackage{dcolumn}
\usepackage{bm}
\usepackage{slashed}
\usepackage{color}
\usepackage[font=small]{caption}
\usepackage{subcaption}
\usepackage{booktabs}

\newcommand{\pTZ}{\langle p_{T,Z}\rangle}

\newcommand{\gosam}{{\tt GoSam}}
\newcommand{\qgraf}{{\tt QGRAF}}
\newcommand{\form}{{\tt FORM}}

\newcommand{\ninja}{{\tt Ninja}}
\newcommand{\spinney}{{\tt Spinney}}
\newcommand{\avholo}{{\tt OneLOop}}

\def\taun{{\cal T}_N}

\def\tauncut{{\cal T}_N^{cut}}

\def\tauzero{{\cal T}_0}
\def\tauzerocut{{\cal T}_0^{cut}}

\title{NNLO predictions for Z-boson pair production at the LHC}
\preprint{\vbox{\hbox{MPP-2017-225}}}
\author[a]{G.~Heinrich,}
\author[a]{S.~Jahn,}
\author[a]{S.~P.~Jones,}
\author[a]{M.~Kerner,}
\author[a]{J.~Pires}
\affiliation[a]{Max Planck Institute for Physics, F\"ohringer Ring 6, 80805 M\"unchen, Germany}

\emailAdd{gudrun@mpp.mpg.de, sjahn@mpp.mpg.de, sjones@mpp.mpg.de, kerner@mpp.mpg.de,pires@mpp.mpg.de}


\abstract{%
 We present a calculation of the NNLO QCD corrections to Z-boson pair
 production at hadron colliders, based on the N-jettiness method for the real radiation parts.
 We discuss the size and shape of the perturbative corrections along with their associated scale uncertainties
 and compare our results to recent LHC data at $\sqrt{s}=13$ TeV.}

\begin{document}

\maketitle

\section{Introduction}\label{sec:intro}

The pair production of $Z$-bosons at the LHC is an important process to test the electroweak sector of the Standard Model (SM). It is sensitive to anomalous gauge boson couplings and constitutes an irreducible background to the production of a Higgs boson decaying into vector bosons and to New Physics searches. 

Recent measurements already include a combined ATLAS and CMS study of anomalous triple gauge couplings in $ZZ$ production based on Run I data~\cite{ATLAS-CONF-2016-036}, as well as measurements at 8\,TeV~\cite{Chatrchyan:2013oev,Khachatryan:2015pba,Aaboud:2016urj} 
and 13\,TeV~\cite{Aad:2015zqe,Khachatryan:2016txa,Aaboud:2017rwm,Sirunyan:2017zjc,ATLAS-CONF-2017-058}.

The NLO QCD corrections to $Z$-boson pair production
were calculated first for stable $Z$-bosons in Refs.~\cite{Ohnemus:1990za,Mele:1990bq}, 
and later including leptonic decays in the
narrow-width approximation in Ref.~\cite{Ohnemus:1994ff}. 
Leptonic decays including spin correlations and off-shell effects have been taken into account in Refs.~\cite{Campbell:1999ah,Dixon:1999di}.

$Z$-boson pair production via gluon fusion is suppressed by two powers of the strong coupling 
compared to the $q\bar{q}$ channel, but contributes significantly to the total cross section due to the large gluon flux at the LHC. The one-loop calculation for stable $Z$-bosons has been performed in Refs.~\cite{Dicus:1987dj,Glover:1988rg}, 
leptonic decays and off-shell effects have been included and studied in Refs.~\cite{Matsuura:1991pj,Zecher:1994kb,Binoth:2008pr,Campbell:2011bn,Kauer:2013qba,Cascioli:2013gfa,Campbell:2013una,Kauer:2015dma}. Soft gluon resummation to the signal/background interference process $gg(\to H^{(*)})\to ZZ$ also has been considered~\cite{Li:2015jva}.

Recently, the 2-loop amplitudes for $q\bar{q}\to VV^{\prime}$~\cite{Gehrmann:2014bfa,Caola:2014lpa,Gehrmann:2015ora} and $gg\to VV^{\prime}$~\cite{vonManteuffel:2015msa,Caola:2015ila} became available, 
and led to the calculation of the NNLO corrections for $Z$-boson pair production, for on-shell $Z$-bosons~\cite{Cascioli:2014yka} as well as including leptonic decays~\cite{Grazzini:2015hta}.
The two-loop corrections to the gluon fusion channel were also calculated~~\cite{Caola:2015psa,Caola:2016trd}
and even combined with a parton shower in Ref.~\cite{Alioli:2016xab}.

Electro-weak (EW) NLO corrections were first calculated for stable vector bosons~\cite{Accomando:2004de,Bierweiler:2013dja,Baglio:2013toa}, and including 
decays within the {\tt Herwig++} framework~\cite{Gieseke:2014gka}. 
Very recently, NLO EW corrections including full off-shell effects have become available~\cite{Biedermann:2017umu,Biedermann:2016yvs,Biedermann:2016lvg}.

The calculation in Refs.~\cite{Cascioli:2014yka,Grazzini:2015hta} is based on the $q_T$-subtraction scheme~\cite{Catani:2007vq} for the doubly unresolved real radiation occurring at NNLO. 
In this letter, we report on the calculation of the NNLO corrections to on-shell $Z$-boson pairs using a different method, 
based on $N$-jettiness subtraction~\cite{Boughezal:2015dva,Gaunt:2015pea}. 
The effect of massive quark loops has been estimated to be at the level of a permille contribution to the total cross section in Ref.~\cite{Cascioli:2014yka}. However,  calculations performed in an $s/m^2_t$ expansion framework~\cite{Melnikov:2015laa,Campbell:2016ivq} indicate that the contributions may be larger, 
and they certainly will be important in the region of large values of the 4-lepton invariant mass $m_{4l}$, which 
is sensitive to the coupling of the longitudinal $Z$-boson components to the top quarks loops.

\section{Details of the calculation}\label{sec:calculation}

The NNLO computation requires the evaluation of the tree-level scattering amplitudes with two additional partons (double-real (RR) contribution), of the one-loop amplitudes with one additional parton (real-virtual (RV) contribution) 
and the two-loop corrections to the Born process (double-virtual (VV) contribution). In this way we systematically combine all the amplitudes containing two additional powers 
in the strong coupling constant with respect to the Born process such that the final result is NNLO accurate in perturbation theory. In Table~\ref{tab:me} we list the matrix elements for $ZZ$ production at NNLO.

\begin{table}
\centering
\label{tab-1}       \begin{tabular}{lll}
\hline
NNLO contributions & perturbative order  \\\hline
$0\to qZZgg\bar{q}$ & tree-level \\
$0\to qZZQ\bar{Q}\bar{q}$ & tree-level  \\
\hline
$0 \to qZZg\bar{q}$   &  one-loop \\
\hline
$0 \to ggZZ     $   &  one-loop \\
$0\to q\bar{q}ZZ$    &  two-loop \\\hline
\end{tabular}
\caption{Perturbative order of the matrix elements for $ZZ$ production at NNLO.}
\label{tab:me}
\end{table}

Although the sum of virtual and real corrections yields a finite
result, the individual contributions contain singularities of infrared (IR) and  ultraviolet nature, 
such that a direct numerical evaluation is not possible. Virtual and real corrections come from phase space integrals 
of different multiplicity; therefore a framework to combine them must be such that the divergent regions in the real-radiation contribution (corresponding to soft and collinear emissions which map to 
configurations with one or two particles less, and therefore are degenerate with the virtual contribution) can be extracted and cancelled with the singularities of the virtual matrix elements.

In this work we employ the $N$-jettiness subtraction scheme~\cite{Gao:2012ja,Boughezal:2015dva,Boughezal:2015aha,Gaunt:2015pea} to perform the evaluation of the NNLO cross section. We begin by reviewing the definition of the $N$-jettiness variable introduced in Refs.~\cite{Stewart:2009yx,Stewart:2010tn},
\begin{equation}
\tau_N = \frac{2}{Q^2} \sum_k \text{min} \left\{q_a \cdot p_k, q_b \cdot p_k,q_1 \cdot p_k,\hdots,q_N \cdot p_k \right\},
\label{eq:TauN}
\end{equation}
where $N$ denotes the number of jets desired in the final state and the sum runs over all QCD  radiated particles. 
In Eq.~\eqref{eq:TauN} the $q_a,q_b$ and $q_1,\hdots,q_N$ are a fixed set of
massless reference momenta for the two beam jets and the $N$ observed jets, the $p_k$ are the parton momenta, and the dimensionful parameter $Q^2$ is the hard interaction scale. For the specific case of a colourless
diboson system in the final state, Eq.~\eqref{eq:TauN} reduces to the $0$-jettiness or beam thrust which in the leptonic frame reads~\cite{Stewart:2009yx,Moult:2016fqy},
\begin{equation}
{\cal T}_0= Q\,\tau_0=\sum_k \text{min} \left\{e^{Y_{ZZ}} n_a \cdot p_k , e^{-Y_{ZZ}} n_b \cdot p_k \right\},
\label{eq:Tau0}
\end{equation}
where $n_a=(1,0,0,1)$ and $n_b=(1,0,0,-1)$ define the beam axis and the $p_k$ are defined in the hadronic centre-of-mass frame. In the context of $N$-jettiness subtractions, taking into
account the boost with rapidity $Y_{ZZ}$ of the Born system ensures that the power corrections are independent of $Y_{ZZ}$~\cite{Moult:2016fqy}.

Looking at the definition of $0$-jettiness in Eq.~\eqref{eq:Tau0} we
can observe that ${\cal T}_0\to 0$ in the limit where the QCD emission $p_k$ is soft or collinear to the initial state. 
For this reason 
values of ${\cal T}_0$ close to zero indicate a final state containing the $ZZ$ pair and only IR (soft and collinear) emissions. In this way the $N$-jettiness variable
can be used as a slicing parameter in any real-radiation phase space
integral to separate infrared singular regions from hard and resolved
configurations. In that sense the approach 
extends the slicing methods developed in the early 90's to compute higher-order corrections at NLO~\cite{Giele:1993dj} to NNLO.

To proceed we employ a ${\cal T}_0^{cut}$ in the real-radiation NNLO phase space and split the cross section into regions above 
and below ${\cal T}_0^{cut}$~\cite{Boughezal:2015dva,Boughezal:2015aha,Gaunt:2015pea},
\begin{equation}
\label{eq:partition}
\begin{split}
\sigma_{NNLO} &= \int {\rm d}\Phi_N \, |{\cal M}_{VV}|^2 +\int {\rm d}\Phi_{N+1} \, |{\cal M}_{RV}|^2 \, \theta_0^{<} \\
&+\int {\rm d}\Phi_{N+2} \, |{\cal M}_{RR}|^2 \, \theta_0^{<}+\int {\rm d}\Phi_{N+1} \, |{\cal M}_{RV}|^2 \, \theta_0^{>} \\
&+\int {\rm d}\Phi_{N+2} \, |{\cal M}_{RR}|^2 \, \theta_0^{>} \\
& \equiv  \sigma_{NNLO}(\tauzero < \tauzerocut)+\sigma_{NNLO}(\tauzero > \tauzerocut)\;.
\end{split}
\end{equation}
In Eq.~(\ref{eq:partition}) we have abbreviated $\theta_0^{<} = \theta(\tauzerocut-\tauzero)$ and $\theta_N^{>} = \theta(\tauzero-\tauzerocut)$, 
and have suppressed any (infrared-safe) measurement function under the phase space integral. The first three terms in this expression all have $\tauzero<\tauzerocut$, and 
are collectively denoted as $\sigma_{NNLO}(\tauzero < \tauzerocut)$, while   
the remaining two terms, with $\tauzero>\tauzerocut$, are collectively denoted as $\sigma_{NNLO}(\tauzero > \tauzerocut)$.  
Contributions with Born-level kinematics necessarily have $\tauzero=0$.\footnote{Prior to its application
for fixed-order perturbative QCD calculations a similar partitioning of the phase space was introduced by the \textsc{Geneva} collaboration~\cite{Alioli:2012fc,Alioli:2013hqa} in the context of merging
fixed-order calculations with parton showers.} 
Contributions with $\tauzero > \tauzerocut$ necessarily contain one or more well separated hadronic energy 
deposits and thus reproduce the $ZZ$+jet cross section at NLO. 
The contributions with $\tauzero < \tauzerocut$ correspond to the limit of the $ZZ$+jet NLO cross section where the jet
is unresolved. The key advantage that allows the computation of the cross section at NNLO below $\tauzerocut$ is the fact that in 
the limit where all QCD emission is soft or collinear, the cross section can be approximately computed using the machinery of Soft-Collinear Effective Theory~(SCET)~\cite{Bauer:2000ew}. In particular,
the existence of a factorization theorem that gives an all-orders description of $N$-jettiness for small $\taun$ less than some value $\tauncut$ allows the cross section to be written in the schematic form,
\begin{equation}
 \sigma(\taun < \tauncut)=  \int H \otimes B \otimes B \otimes S \otimes   \left[ \prod_{n}^{N} J_n \right] +\cdots ,
\label{eq:SCET}
\end{equation}
where $H$ describes the effect of hard radiation from the purely virtual corrections to the process, $B$ encodes the effect of radiation collinear to one of the two
initial beam directions, $S$ describes soft radiation and $J_n$ contains the radiation collinear to hard final-state jets. The ellipsis denote power-suppressed terms
which become negligible for $\taun \ll Q$.

We have expanded the formula in Eq.~\eqref{eq:SCET} to second order in the strong coupling constant to obtain the $\sigma_{NNLO}(\tauzero < \tauzerocut)$ cross section for $ZZ$ production
at hadron colliders. In particular this includes contributions from the universal quark beam function at two loops~\cite{Gaunt:2014xga} and the $0$-jettiness soft function 
at two-loops~\cite{Kelley:2011ng,Monni:2011gb}. The process dependent hard function has been extracted
from the two-loop amplitude computed in Ref.~\cite{Gehrmann:2015ora} via an interface to the program \texttt{qqvvamp}.
We do not include massive top-quark loops in the $q\bar{q}ZZ$ two-loop amplitude. Using $N_f=5$ therefore introduces the chiral anomaly stemming from subdiagrams 
where one $Z$-boson and two gluons are attached to a $b$-quark triangle. However, we neglect this anomalous contribution in our calculation as the anomaly must cancel 
once the top quark loops are included, following the same strategy as advocated in Ref.~\cite{Caola:2015psa}.

In SCET renormalised form, the IR finite one- and two-loop amplitudes can be written at renormalisation scale $\mu^2$ as~\cite{Becher:2009cu,Becher:2009qa}
\begin{eqnarray}
{\Omega}_{N}^{(1),\textrm{finite}}&=&{\Omega}^{(1)}-I_{1}^{N}(\epsilon){\Omega}^{(0)},\nonumber\\
{\Omega}_{N}^{(2),\textrm{finite}}&=&{\Omega}^{(2)}-I_{1}^{N}(\epsilon){\Omega}^{(1)}-I_{2}^{N}(\epsilon){\Omega}^{(0)},
\label{eq:SCETpoles}
\end{eqnarray}
where the $N$-jettiness subtraction operators are defined
by\footnote{The subtraction operators are identical to those given for 
diphoton production in Appendix A, Eq.~(A.17) of Ref.~\cite{Campbell:2016yrh}, 
but these formulae appear to contain two typographical errors. 
Specifically, we find that $\Gamma_0^\prime$ should be multiplied to $\beta_0$ 
in the second term of Eq.~(\ref{subtraction_operators}) and that the last term has a 
factor 16 in the denominator.}
\begin{eqnarray}
I_1^N(\epsilon)&=&\frac{\Gamma_0^\prime}{8\epsilon^2} + \frac{\mathbf{\Gamma}_0}{4 \epsilon},\nonumber\\
I_2^N(\epsilon)&=&-\frac{(\Gamma_0^\prime)^2}{128 \epsilon^4} - \frac{6 \beta_0 \Gamma_0^\prime + 2 \Gamma_0^\prime \mathbf{\Gamma}_0}{64 \epsilon^3} - \frac{8 \beta_0 \mathbf{\Gamma}_0 + 2 (\mathbf{\Gamma}_0)^2 - \Gamma_1^\prime}{64 \epsilon^2} + \frac{\mathbf{\Gamma}_1}{16 \epsilon},
\label{subtraction_operators}
\end{eqnarray}
with
\begin{flalign}
&\Gamma_0^\prime = - 2 C_F \gamma_0^\mathrm{cusp},\quad
\Gamma_1^\prime = - 2 C_F \gamma_1^\mathrm{cusp},\nonumber\\
&\mathbf{\Gamma}_0 = - C_F \gamma_0^\mathrm{cusp} \log \left( \frac{\mu^2}{-s} \right) + 2 \gamma_0^q,\quad
\mathbf{\Gamma}_1 = - C_F \gamma_1^\mathrm{cusp} \log \left( \frac{\mu^2}{-s} \right) + 2 \gamma_1^q,
\end{flalign}
and constants
\begin{flalign}
&\gamma_0^\mathrm{cusp} = 4,\quad
\gamma_1^\mathrm{cusp} = \left(\frac{268}{9} - \frac{4 \pi^2}{3}\right) C_A - \frac{80}{9} T_F n_f,\quad
\gamma_0^q = -3 C_F,\nonumber\\
&\gamma_1^q = 
\left( -\frac{3}{2} + 2 \pi^2 - 24 \zeta_3 \right) C_F^2 + 
\left(- \frac{961}{54} - \frac{11 \pi^2}{6} + 26 \zeta_3 \right) C_F C_A + 
\left( \frac{130}{27} + \frac{2 \pi^2}{3} \right) C_F T_F n_f,\nonumber\\
&\beta_0 = \frac{11}{6} C_A - \frac{4}{6} T_F n_f,\quad
C_A = N,\quad 
C_F = \frac{N^2-1}{2 N},\quad
T_F = \frac{1}{2}.
\end{flalign}

In Ref.~\cite{Gehrmann:2015ora} the finite remainder of the one- and two-loop form factors for vector boson pair production are provided in the 
$q_T$-~\cite{Catani:2013tia} and Catani-~\cite{Catani:1998bh} subtraction schemes. 
By comparing the definition of the subtraction schemes, the form factors in the $N$-jettiness scheme can be derived from those of the $q_T$-scheme according to,
\begin{eqnarray}
{\Omega}_{N}^{(1),\textrm{finite}}&=&{\Omega}_{q_T}^{(1),\textrm{finite}}+\Delta I_{1}{\Omega}_{q_T}^{(0),\textrm{finite}},\nonumber\\
{\Omega}_{N}^{(2),\textrm{finite}}&=&{\Omega}_{q_T}^{(2),\textrm{finite}}+\Delta I_{1}{\Omega}_{q_T}^{(1),\textrm{finite}}+\Delta I_{2}{\Omega}_{q_T}^{(0),\textrm{finite}},\nonumber
\end{eqnarray}
with the scheme conversion coefficients given by
\begin{eqnarray}
\Delta I_1 &=& \frac{\pi^2}{12} C_F,\nonumber\\
\Delta I_2 &=& 
\frac{\pi^4}{288} C_F^2 
+ \left( -\frac{607}{162} +\frac{67 \pi^2}{144} + \frac{77 \zeta_3}{36}-\frac{\pi^4}{72} + \frac{11 i \pi^3}{72} \right) C_A C_F \nonumber\\
&& + \left( \frac{41}{81} -\frac{5 \pi^2}{72} -\frac{7 \zeta_3}{18}  -\frac{i \pi^3}{36} \right) C_F n_f,
\end{eqnarray}
where, for brevity, we have set $\mu^2=s$.

Finally we have obtained the $\sigma_{NNLO}(\tauzero > \tauzerocut)$ contribution of the $ZZ$ NNLO cross section using the
tree level matrix elements from VBFNLO~\cite{Arnold:2011wj,Campanario:2014ioa} for the double-real emission phase space integral cross-checked with {\tt MadGraph5}~\cite{Alwall:2014hca},
while the one-loop  amplitudes for the real-virtual phase space were generated with \gosam{}~\cite{Cullen:2011ac,Cullen:2014yla} and cross-checked with OpenLoops~\cite{Cascioli:2011va}. \gosam{}
uses \qgraf{}~\cite{Nogueira:1991ex}, \form~\cite{Kuipers:2012rf} and
\spinney{}~\cite{Cullen:2010jv} for the generation of the Feynman
diagrams, and offers a choice from {\tt
Samurai}~\cite{Mastrolia:2010nb}, {\tt
golem95C}~\cite{Binoth:2008uq,Cullen:2011kv,Guillet:2013msa}
and \ninja{}~\cite{vanDeurzen:2013saa,Peraro:2014cba} for the
reduction.  At run time the amplitudes were computed using
\ninja{}, which calls \avholo{}~\cite{vanHameren:2010cp} for the master integrals, and rescued using 
an implementation of  \ninja{} in quadruple precision for unstable phase space points. We also include the loop 
induced one-loop squared corrections in the $gg\to ZZ$ channel,  which are formally of NNLO accuracy, keeping full dependence on the top quark mass and on the Higgs mediated contributions using \gosam{}.

\subsection{Discussion of the IR subtraction procedure}
Before we present our numerical results for $ZZ$ production at hadron colliders 
we would like to make a few remarks concerning the IR subtraction scheme employed for this calculation.
As mentioned in the previous section, the $N$-jettiness subtraction scheme is a non-local subtraction scheme. In local subtraction schemes
the IR divergent phase space integrals are regulated by the introduction of suitable IR real-radiation counterterms
that satisfy two basic properties,
\begin{itemize}
\item reproduce locally, for each phase space point in a singular region, the physical IR divergent soft and collinear limits of the matrix elements of the process under consideration;
\item be simple enough to allow their analytic integration and generate a local and pointwise analytic pole cancellation between 
the explicit $1/\epsilon$-poles of the virtual corrections and the $1/\epsilon$-poles of the integrated real-radiation counterterms.
\end{itemize}

Some flexibility however exists on how the singular limits are locally subtracted and how the real-radiation phase space is parametrised.
For antenna subtraction~\cite{GehrmannDeRidder:2005cm,Glover:2010im,Currie:2013vh} 
physical matrix elements with three partons~\cite{Daleo:2006xa} at tree-level and one-loop suffice to reproduce single unresolved
limits in QCD amplitudes, while four parton antennae~\cite{Daleo:2009yj,GehrmannDeRidder:2012ja} subtract doubly 
unresolved configurations at NNLO~\cite{Ridder:2016nkl,Chen:2016zka,Currie:2016bfm,Currie:2017eqf}. 
Examples of local IR subtraction schemes which employ a structured 
decomposition of the real-radiation phase space based on singular IR limits of QCD amplitudes~\cite{Czakon:2010td,Boughezal:2013uia,Czakon:2014oma,Caola:2017dug} have also
been developed and applied to specific NNLO calculations. For the specific case of colourless systems in the initial state local subtractions
have been developed in Refs.~\cite{DelDuca:2015zqa,Somogyi:2017bui}.
The extension of the $N$-jettiness method towards local subtractions has also been conceptually discussed in Ref.~\cite{Gaunt:2015pea}.

On the other hand, non-local IR subtraction schemes regulate the singularities of the real-radiation phase space integrals by the introduction of a 
suitable variable ($N$-jettiness or the transverse momentum $q_T$ of a colourless system for $q_T$-subtraction~\cite{Catani:2007vq}\footnote{An extension of $q_T$-subtraction to colourful final states has been worked out in Ref.~\cite{Bonciani:2015sha}.}) that regulates the phase space integration
by separating IR divergent regions from hard and resolved ones according to Eq.~\eqref{eq:partition}. 
In this way contributions to the cross section
for ${\cal T}_0$ above and below ${\cal T}_0^{cut}$ are separately logarithmically divergent. However, in the sum all
the logarithmic dependence on ${\cal T}_0^{cut}$ should cancel, provided the value of ${\cal T}_0^{cut}$ employed in the phase space integration
is small enough such that the SCET approximation to the cross section is valid. In particular, for each $1/\epsilon$-IR pole in dimensional regularisation
there is a corresponding logarithmically divergent coefficient predicted from SCET in the ${\cal T}_0\to 0$ limit, according to
\begin{equation}
\frac{1}{\epsilon^n}\sim \log^{n}\left(\frac{{\cal T}_0}{\mu} \right).
\end{equation}
The infrared pole cancellation in this case is observed through the cancellation between the universal and analytically known terms predicted by SCET,
integrated over the Born phase space, and the Monte Carlo  integration over the real-radiation phase space of the real-emisson 
matrix elements for small ${\cal T}_0$.
The method of $N$-jettiness meanwhile has been applied successfully to various
processes calculated at NNLO~\cite{Boughezal:2015ded,Boughezal:2015aha,Gao:2012ja,Boughezal:2015dva,Campbell:2016jau,Boughezal:2016dtm,Boughezal:2016isb,Campbell:2016yrh,Boughezal:2016wmq,Campbell:2016lzl,Campbell:2017dqk}.

Due to the non-local IR subtraction method employed in our calculation of the NNLO corrections we found it necessary to do the following optimisations at the Monte Carlo integration
level in order to observe the independence of our results on the choice of the slicing parameter value ${\cal T}_0^{cut}$:
\begin{itemize}
\item introduce a phase space generator where the 0-jettiness variable ${\cal T}_0$ is directly sampled by VEGAS. This can be achieved by applying the transformation
\begin{equation}
p_{\pm,i} = E_i\pm p_{z,i}, \quad \mathrm{d}E_i\, \mathrm{d}p_{z,i} = \frac{1}{2} \mathrm{d}p_{+,i}\,\mathrm{d}p_{-,i}
\end{equation}
to the integration over the real radiation phase space.\footnote{For each transformation, the integration boundaries, which we do not state explicitly,  have to be changed accordingly.}
With the momenta defined in the center-of-mass system of the $Z$-bosons, the 0-jettiness as defined in Eq.~\eqref{eq:Tau0} then reads
\begin{equation}
\mathcal T_0 =  \min(p_{+,1},p_{-,1}) + \min(p_{+,2},p_{-,2}). 
\end{equation}
With further transformations, where the regions with $p_{+,i} <p_{-,i}$ and $p_{+,i} >p_{-,i}$ have to be distinguished, 
it is then possible to directly sample $\mathcal T_0$, followed by the sampling of $\min(p_{+,1},p_{-,1})$, which also fixes the value of $\min(p_{+,2},p_{-,2})$.
Afterwards, the two missing values of $p_{\pm,i}$ can be sampled.

\item have a fast implementation of the double-real and real-virtual matrix elements for $ZZ$-production which is stable in the multiple soft and collinear limits.
\end{itemize}

The first optimisation ensures that the real-radiation phase space generator properly samples the phase space boundaries determined by the choice of slicing parameter ${\cal T}_0^{cut}$
and that the phase space integral converges. 
The second optimisation ensures that the matrix elements are fast and stable enough to be integrated near the singular IR limits
which give the bulk of the cross section for the $(\tauzero > \tauzerocut)$ phase space integrals when $\tauzerocut$ is small. This is in contrast with a local subtraction scheme, where the 
real radiation subtraction terms ensure that the integrand vanishes as we approach a singular region.

\section{Results}\label{sec:results}

Our numerical studies for proton-proton collisions at centre-of-mass energy $\sqrt{s}=$13 TeV are for on-shell $Z$-boson pair production.
We use the MSTW2008~\cite{Martin:2007bv} and NNPDF-3.0~\cite{Ball:2014uwa} sets of parton distribution functions via the LHAPDF~\cite{Buckley:2014ana} interface, with densities
and $\alpha_s$ evaluated at each corresponding order (i.e. we use ($n$+1)-loop
$\alpha_s$ at N$^{n}$LO, with $n=0,1,2$) and we consider $N_f=5$ massless quark flavours. The default renormalisation ($\mu_R$) and factorisation ($\mu_F$) scales
are set to $\mu_R=\mu_F=m_{Z}$. We use the $G_\mu$ EW scheme where the EW input parameters have been set to $G_F=1.16639\times 10^{-5}$, $m_W=80.399$ GeV and
$m_Z=91.1876$ GeV. The top quark and Higgs boson masses that are included in the RV one-loop contributions and in the loop-induced $gg$ channel have been set to $m_t=173.2$\,GeV and
$m_H=125$\,GeV, respectively. We do not include top quark contributions in the double virtual two-loop diagrams.

\begin{figure}
\centering
\begin{subfigure}{.5\textwidth}
  \centering
  \includegraphics[width=3.2in]{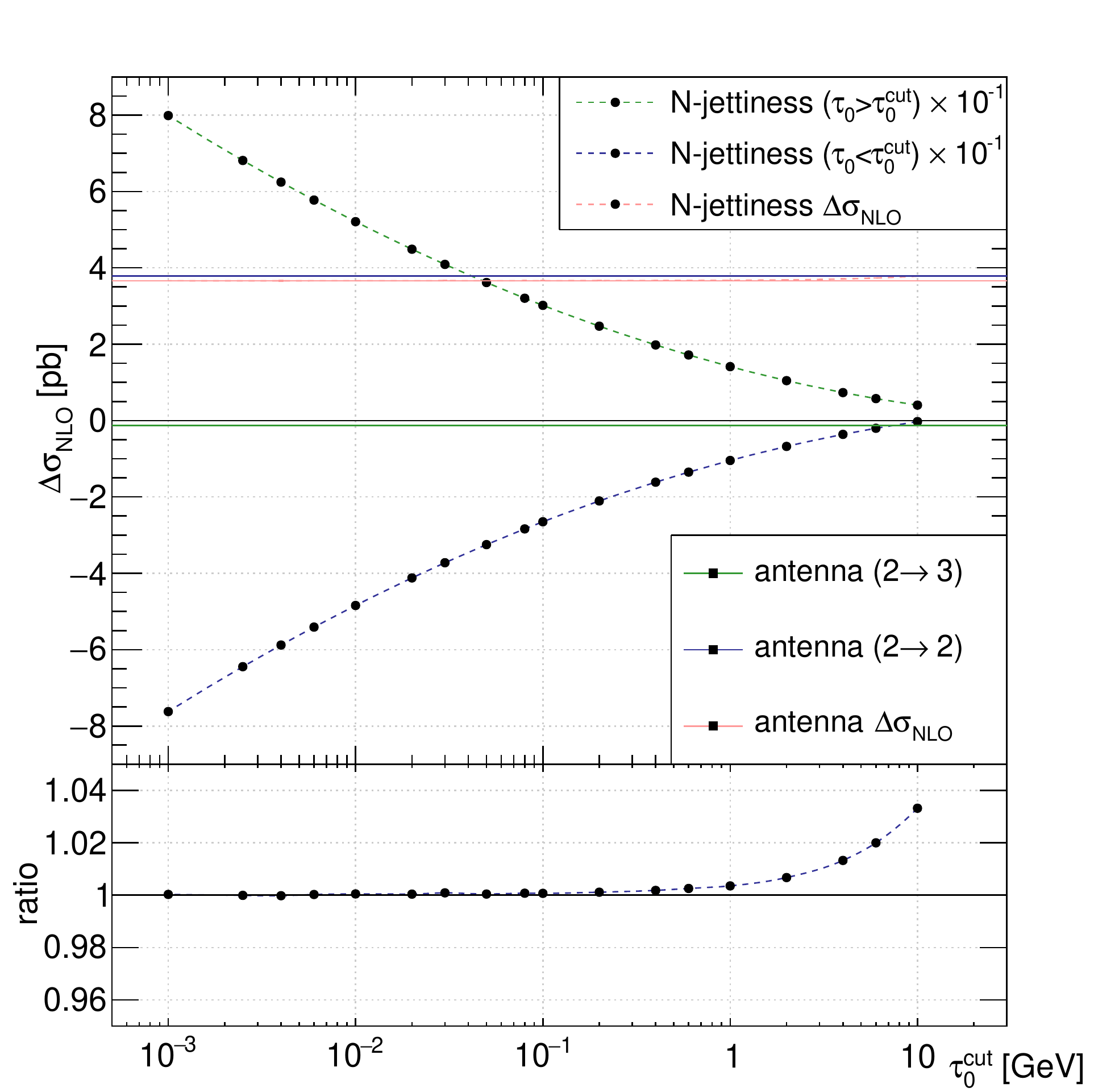}
  \caption{}
  \label{fig:tau1sub1}
\end{subfigure}%
\begin{subfigure}{.5\textwidth}
  \centering
  \includegraphics[width=3.5in]{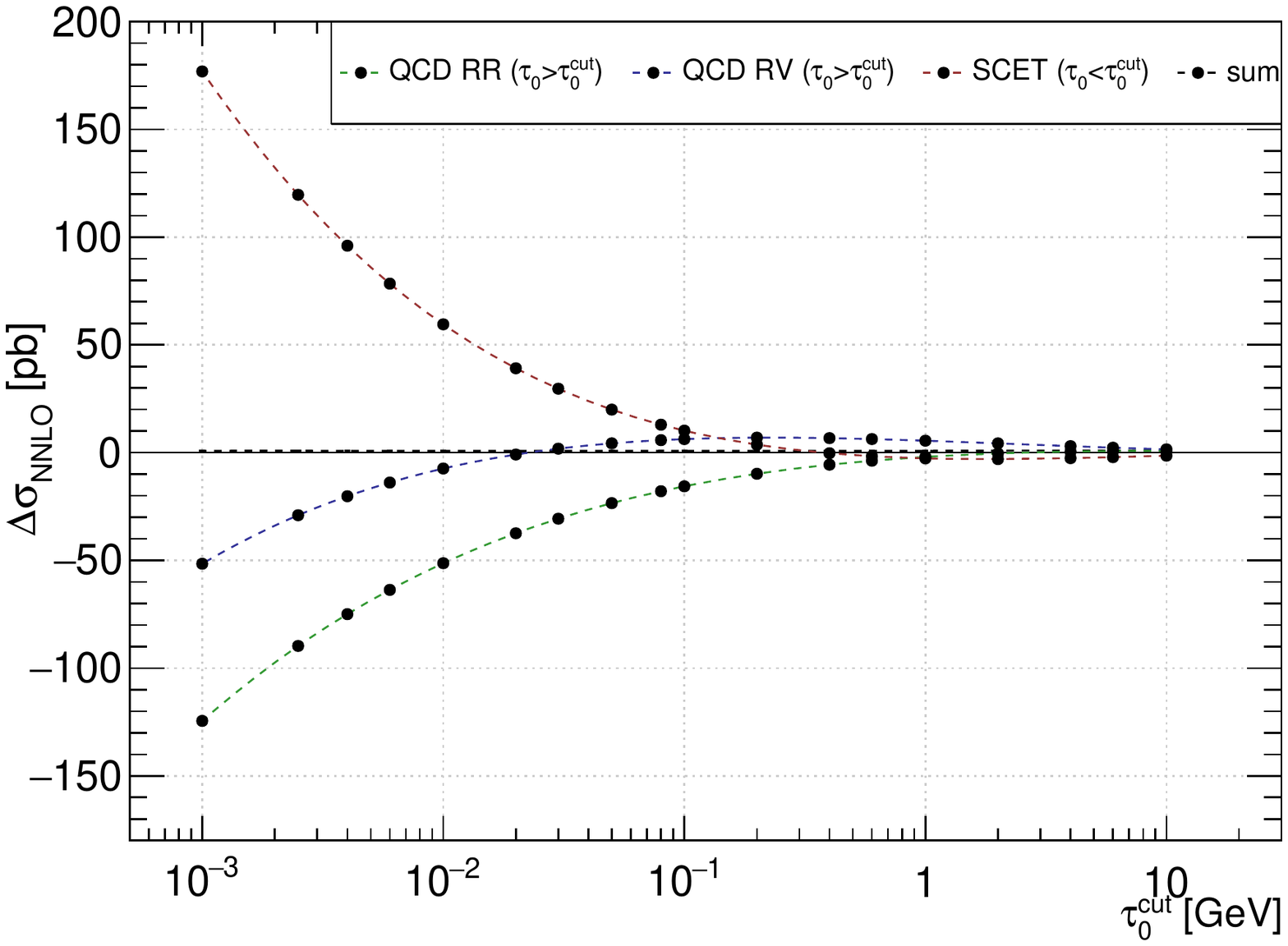}
  \caption{}
  \label{fig:tau1sub2}
\end{subfigure}
\caption{(a) NLO coefficient to the inclusive $ZZ$ cross section computed with $N$-jettiness subtraction (dashed-lines) and antenna subtraction (solid lines) as a function of ${\cal T}_0^{cut}$. 
 We show separately the contributions of the $2\to3$ and $2\to2$ phase space integrals for antenna subtraction, and 
  the  $\sigma_{NLO}(\tauzero > \tauzerocut)$  and $\sigma_{NLO}(\tauzero < \tauzerocut)$ phase space integrals for $N$-jettiness. 
  The ratio plot shows $\Delta\sigma_{NLO}$($N$-jettiness) over $\Delta\sigma_{NLO}$(antenna).
  (b) NNLO coefficient to the inclusive $ZZ$ cross 
  section computed with $N$-jettiness subtraction (dashed lines) as a function of ${\cal T}_0^{cut}$. We show the separate cross sections for $\sigma_{NNLO}(\tauzero > \tauzerocut)$  from the double-real and real-virtual
  phase space integrals and for $\sigma_{NNLO}(\tauzero < \tauzerocut)$ from the SCET phase space integrals together with their sum.}
\label{fig:tau1}
\end{figure}

We show in Fig.~\ref{fig:tau1} the NLO and NNLO coefficients of the $ZZ$ cross section as a function of ${\cal T}_0^{cut}$. On the left-hand side
we compare the NLO results obtained using either antenna subtraction or $N$-jettiness subtraction
and observe full agreement in the evaluation of the NLO corrections $\Delta\sigma_{\rm{NLO}}$.  
We present separate results for the
phase space integration with real emission kinematics ($2\to3$) and Born-like kinematics ($2\to2$). Obviously, using a local
subtraction scheme (antenna subtraction), all phase space integrals contributing to $\Delta\sigma_{\rm{NLO}}$ are independent of the choice of ${\cal T}_0^{cut}$. 
Moreover, the bulk of the NLO coefficient comes from
the $2\to2$ phase space integral which determines where more statistical precision is needed to obtain an accurate result. On the other hand, using $N$-jettiness we observe 
that both phase space integrals are separately double-logarithmically divergent and therefore a very good numerical precision is needed for both contributions to improve the accuracy
of the final result. In Fig.~\ref{fig:tau1sub2} we present the NNLO coefficient of the $ZZ$ cross section as a function of ${\cal T}_0^{cut}$. In this case we observe
that the phase space integrals for the contributions $\sigma_{NNLO}(\tauzero < \tauzerocut)$ and $\sigma_{NNLO}(\tauzero > \tauzerocut)$ are logarithmically divergent 
to the fourth power in $\log\left({\cal T}_0^{cut}\right)$, and 
for typical values of ${\cal T}_0^{cut}$ in the range $10^{-2} - 10^{-3}$\,GeV need to be known with better than permille level accuracy to achieve an accurate
determination of the NNLO coefficient. 
  
\begin{figure}[t]
\centering
\includegraphics[width=6in]{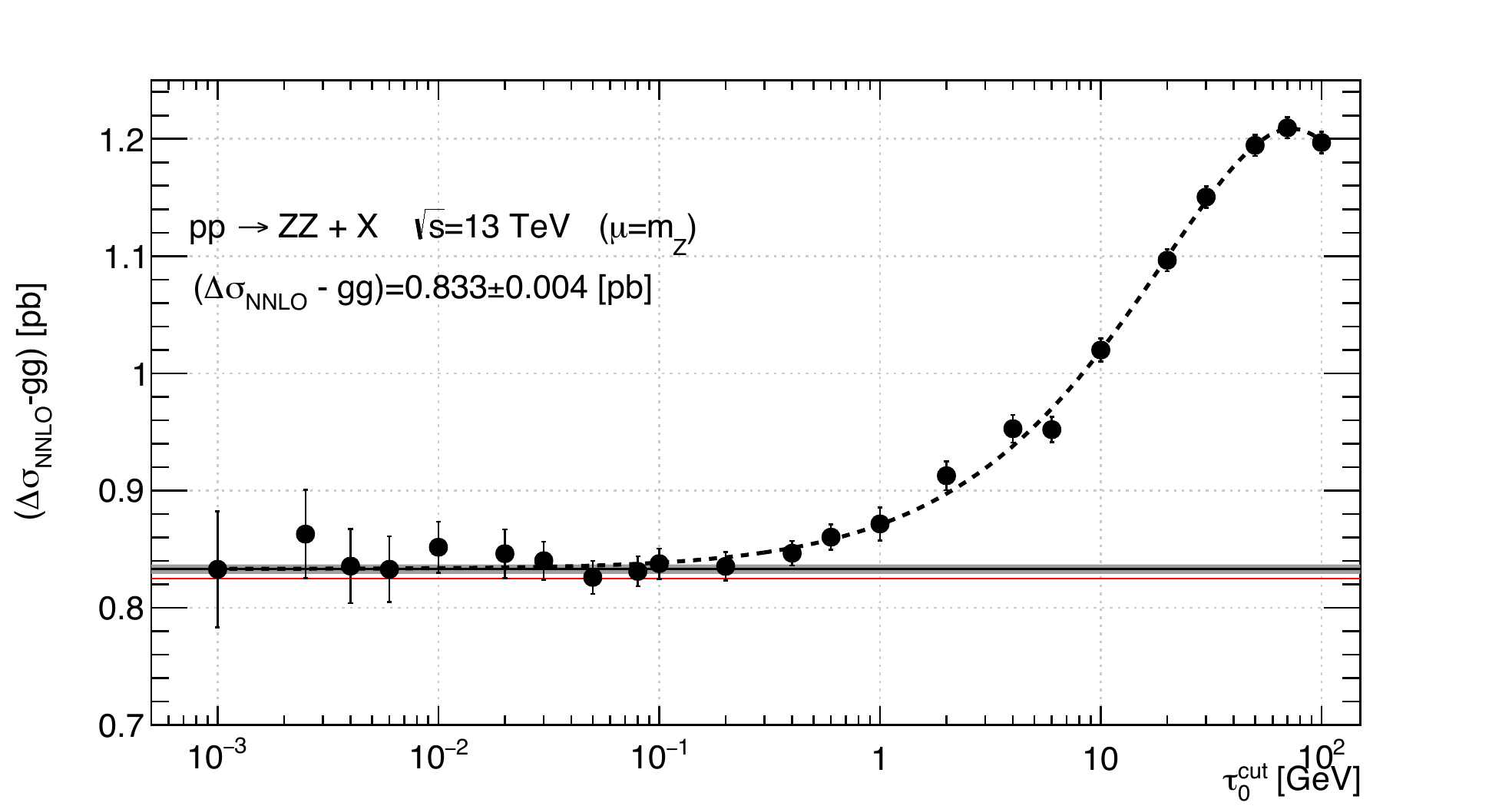}
\caption{$\tauzerocut$ dependence of the NNLO coefficient for $ZZ$ production with the $\tauzero$ independent $gg\to ZZ$ contribution subtracted. The black dashed line shows the fit of the $\tauzerocut$ dependence of 
  the NNLO coefficient (black data points) to the analytic form in equation~\eqref{eq:powercorr}. 
The $\tauzerocut \rightarrow 0$ limit is shown as a solid black line with a grey band showing the uncertainty on the fitted parameter.  The red line represents
  the NNLO coefficient reconstructed from the NNLO result obtained in Ref.~\cite{Cascioli:2014yka}.  }
\label{fig:tau2}
\end{figure}

In order to study in more detail the independence of the NNLO coefficient on the choice of slicing parameter ${\cal T}_0^{cut}$ we present in Fig.~\ref{fig:tau2} 
on a smaller scale the NNLO coefficient after combining the contributions for $\sigma_{NNLO}(\tauzero < \tauzerocut)$ and $\sigma_{NNLO}(\tauzero > \tauzerocut)$ as 
black data points. Within the errors we observe a plateau in the region ${\cal T}_0^{cut}=10^{-1}\sim10^{-3}$ GeV where the results tend to a constant. In addition we can observe for larger values of ${\cal T}_0^{cut}$ (${\cal T}_0^{cut} >10^{-1}$ GeV) the on-set of the power corrections to the $N$-jettiness SCET 
factorisation theorem which we do not compute. The
fact that the on-set of power corrections shows up for fairly large values of ${\cal T}_0^{cut}$ with respect to other processes~\cite{Boughezal:2015dva,Boughezal:2015aha,Boughezal:2016wmq} seems to indicate that
for $ZZ$ production their contribution is small. Nonetheless the leading power correction can be modeled after integration over the final-state phase space as~\cite{Moult:2016fqy,Boughezal:2016zws,Moult:2017jsg}

\begin{equation}
\Delta\sigma_{jettiness}^{NNLO}(\tauzerocut)=\Delta\sigma^{NNLO}+c_3\,\frac{\tauzerocut}{Q}\,\log^3\left(\frac{\tauzerocut}{Q}\right)
+c_2\,\frac{\tauzerocut}{Q}\,\log^2\left(\frac{\tauzerocut}{Q}\right)+\hdots,
\label{eq:powercorr}
\end{equation}
where $Q$ is an appropriate hard scale of the process and $c_2, c_3$ are unknown constants. 
We have performed a fit of the results of our Monte-Carlo runs to this functional form of the $N$-jettiness NNLO coefficient for $ZZ$ and show the resulting fit as
a black dotted line in Fig.~\ref{fig:tau2}. The fit allows us to numerically extract the value of the NNLO coefficient in the limit where $\tauzero\to0$. This value can be compared 
to the reconstructed NNLO coefficient for $ZZ$ production obtained in Ref.~\cite{Cascioli:2014yka}\footnote{The
  NNLO coefficient was reconstructed by subtracting from the total NNLO $ZZ$ cross section quoted in Table 1 of Ref.~\cite{Cascioli:2014yka} the NLO $ZZ$ cross section
and the contribution from the loop-induced $gg$-channel.}, which is shown as a red line. 
We use the extrapolated value for our result for the $ZZ$ cross section at NNLO shown in Table~\ref{tab:Xsec}, which is in excellent agreement with the result $\sigma_{NNLO} = 16.91\,\mathrm{pb}$ obtained in Ref.~\cite{Cascioli:2014yka}. 

As a consistency check we have also fitted a constant to the plateau region 
($\tauzerocut < 10^{-2}$ GeV or $\tauzerocut < 10^{-1}$ GeV) and  
these fits yield compatible results for $\Delta\sigma^{NNLO}$. Further, 
we have also fitted the leading power corrections using~\eqref{eq:powercorr} 
including only results for $\tauzerocut < 1$ GeV. When fitting the 
leading power corrections with $\tauzerocut < 1$ GeV there is a strong 
correlation between $c_3$ and $Q$ as well as $c_2$ and $Q$; fixing $Q$ 
to values in the range $50-5000$ GeV we obtain compatible results for 
$\Delta\sigma^{NNLO}$. Including in the fit results up to 
$\tauzerocut < 10^2$ GeV, as shown in Fig.~\ref{fig:tau2} , we obtain 
a stable fit also when $Q$ is treated as a free parameter.

\begin{table}[]
\centering
\begin{tabular}{|c|c|c|c|}
\hline
  & $\sigma_{LO}$ [pb] & $\sigma_{NLO}$ [pb] & $\sigma_{NNLO}$ [pb]\\ \hline
Our Result & & & \\  
MSWT2008 &  $9.890^{+4.9\%}_{-6.1\%}  $   & $14.508^{+3.0\%}_{-2.4\%}$    &    $16.92^{+3.2\%}_{-2.6\%} $  \\ 
NNPDF3.0 &  $9.845^{+5.2\%}_{-6.2\%}  $   & $14.100^{+2.9\%}_{-2.4\%}$    &    $16.69^{+3.1\%}_{-2.8\%} $  \\ \hline
ATLAS~\cite{Aaboud:2017rwm} & \multicolumn{3}{c|}{$17.3 \pm 0.6 (\mathrm{stat.}) \pm 0.5 (\mathrm{syst.}) \pm 0.6 (\mathrm{lumi.}) $} \\ \hline
CMS~\cite{Sirunyan:2017zjc} & \multicolumn{3}{c|}{$17.2 \pm 0.5 (\mathrm{stat.}) \pm 0.7 (\mathrm{syst.}) \pm 0.4 (\mathrm{theo.}) \pm 0.4 (\mathrm{lumi.})$} \\ \hline
\end{tabular}
\caption{Inclusive cross section for $ZZ$ production at the LHC run II $\sqrt{s}=$13 TeV at LO, NLO and NNLO with $\mu_R=\mu_F=m_{Z}$,
together with the measurements from ATLAS~\cite{Aaboud:2017rwm} and CMS~\cite{Sirunyan:2017zjc}. Uncertainties in the theory calculation at each order are obtained
by varying the renormalisation and factorisation scales in the range $0.5m_{Z}<\mu_R,\mu_F<2m_{Z}$ with the constraint 
$0.5<\mu_F/\mu_R<2$. Uncertainties in the experimental measurements denote absolute statistical, systematic and luminosity uncertainties.}
\label{tab:Xsec}
\end{table}

The resulting theoretical predictions can be compared with the ATLAS and CMS measurements at $\sqrt{s}=13$ TeV~\cite{Aaboud:2017rwm,Sirunyan:2017zjc}, also shown in Table~\ref{tab:Xsec}. In the same Table we present an updated value for the NNLO cross section computed as described in the previous section
using the more recently determined NNPDF-3.0~\cite{Ball:2014uwa} PDF sets and an updated value for the $W$-boson mass of $M_W$=80.385\,GeV;
these settings are also used for our phenomenological results in the following.
We observe a significant improvement in the agreement with the data after the inclusion of the NNLO corrections. 

\begin{figure}[t]
\centering
\includegraphics[width=4.5in]{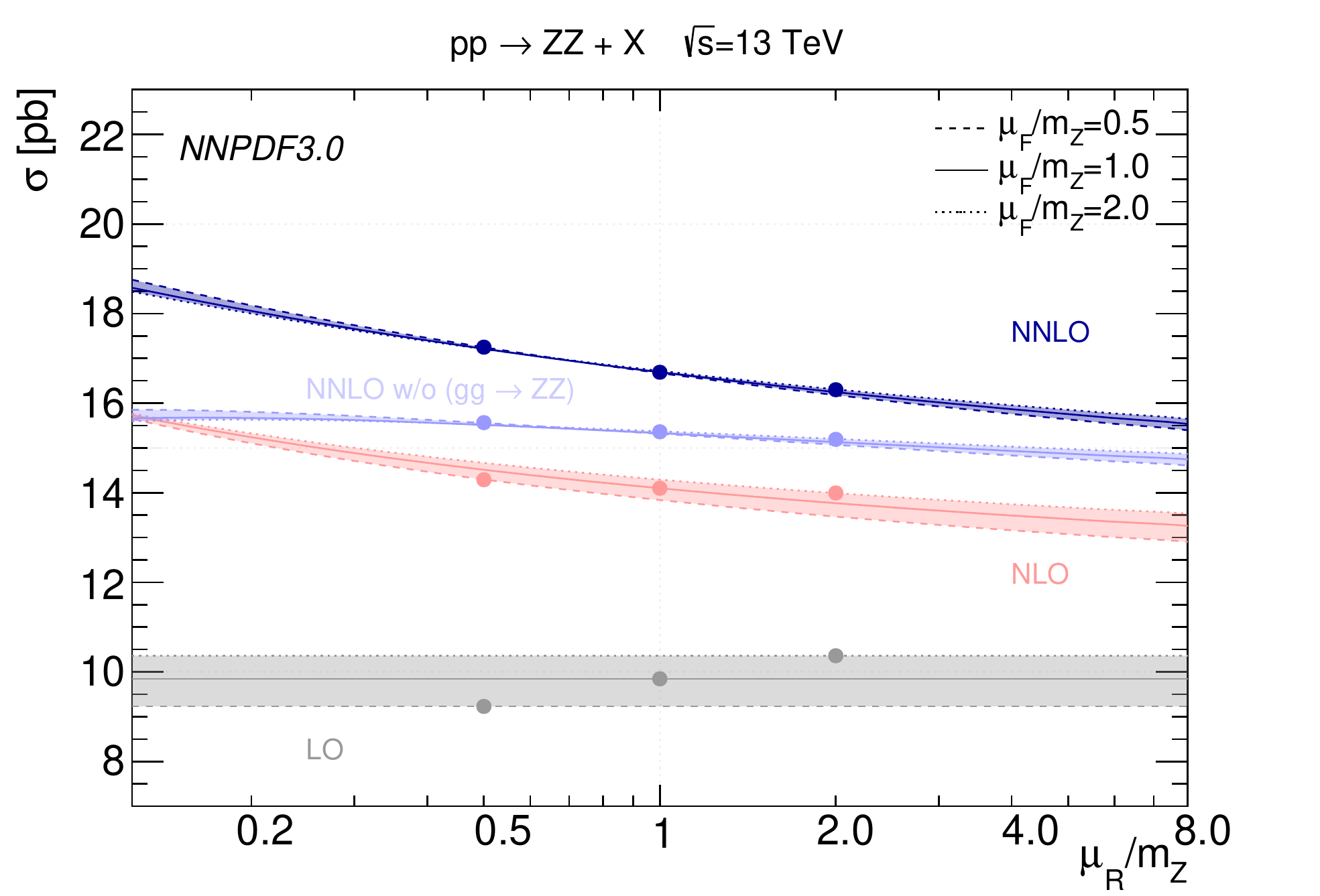}
\caption{Renormalisation and factorisation scale dependence of the $ZZ$ cross section at LO, NLO and NNLO for the central scale choice $\mu_R=\mu_F=m_Z$ and with NNPDF-3.0 PDFs. 
We also show the NNLO result without the gluon fusion contributions.
The thickness of the bands shows the variation in the cross section due to factorisation scale while the slope shows the renormalisation scale dependence. The scale
uncertainty was obtained by varying the renormalisation and factorisation scales in the range $0.5m_{Z}<\mu_R,\mu_F<2m_{Z}$ with the constraint 
$0.5<\mu_F/\mu_R<2$.} 
\label{fig:scale}
\end{figure}

\begin{figure}[t]
\centering
\begin{subfigure}{.5\textwidth}
  \centering
  \includegraphics[width=3.1in]{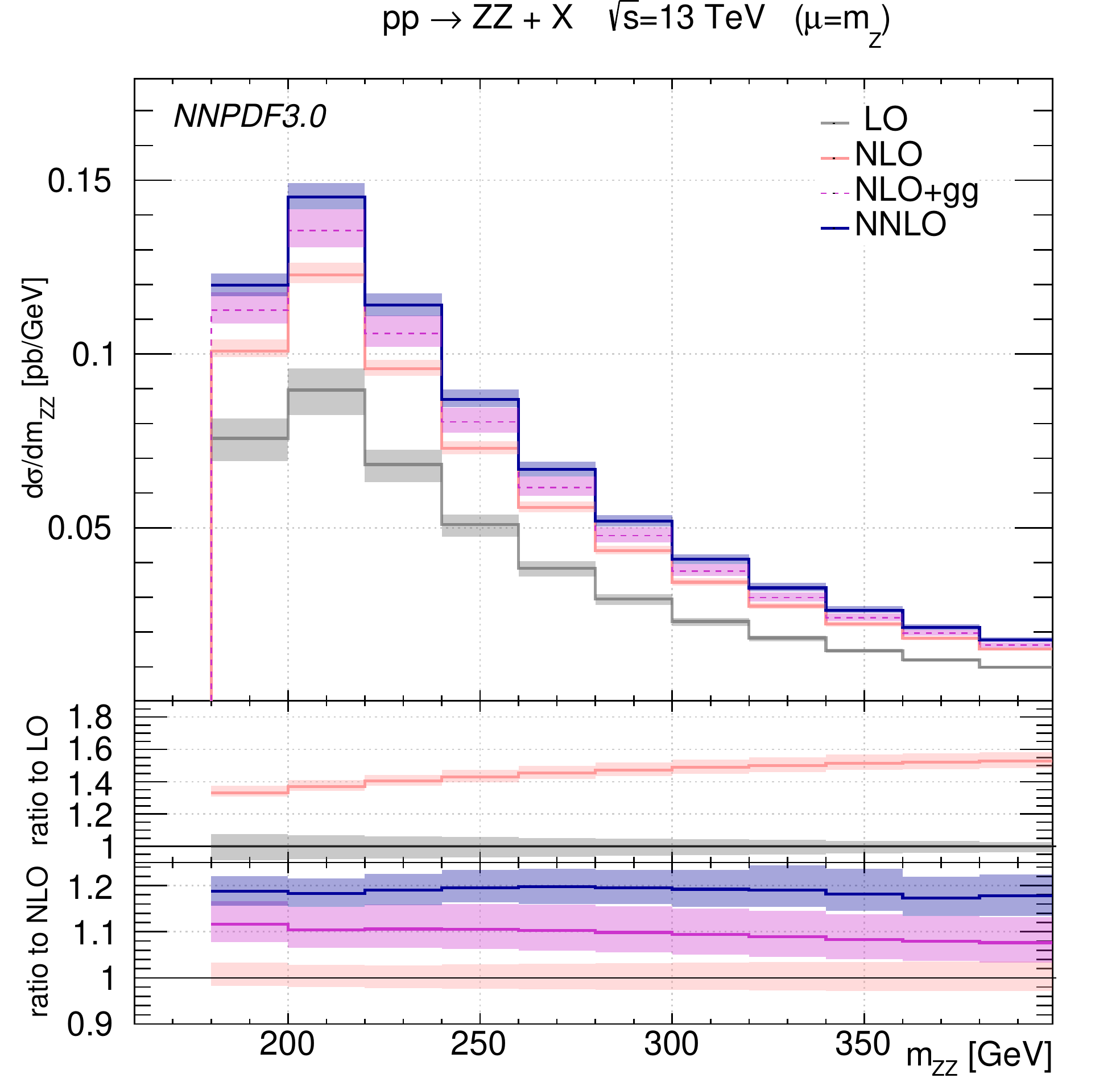}
  \caption{}
  \label{fig:diffxssub1}
\end{subfigure}%
\begin{subfigure}{.5\textwidth}
  \centering
  \includegraphics[width=3.1in]{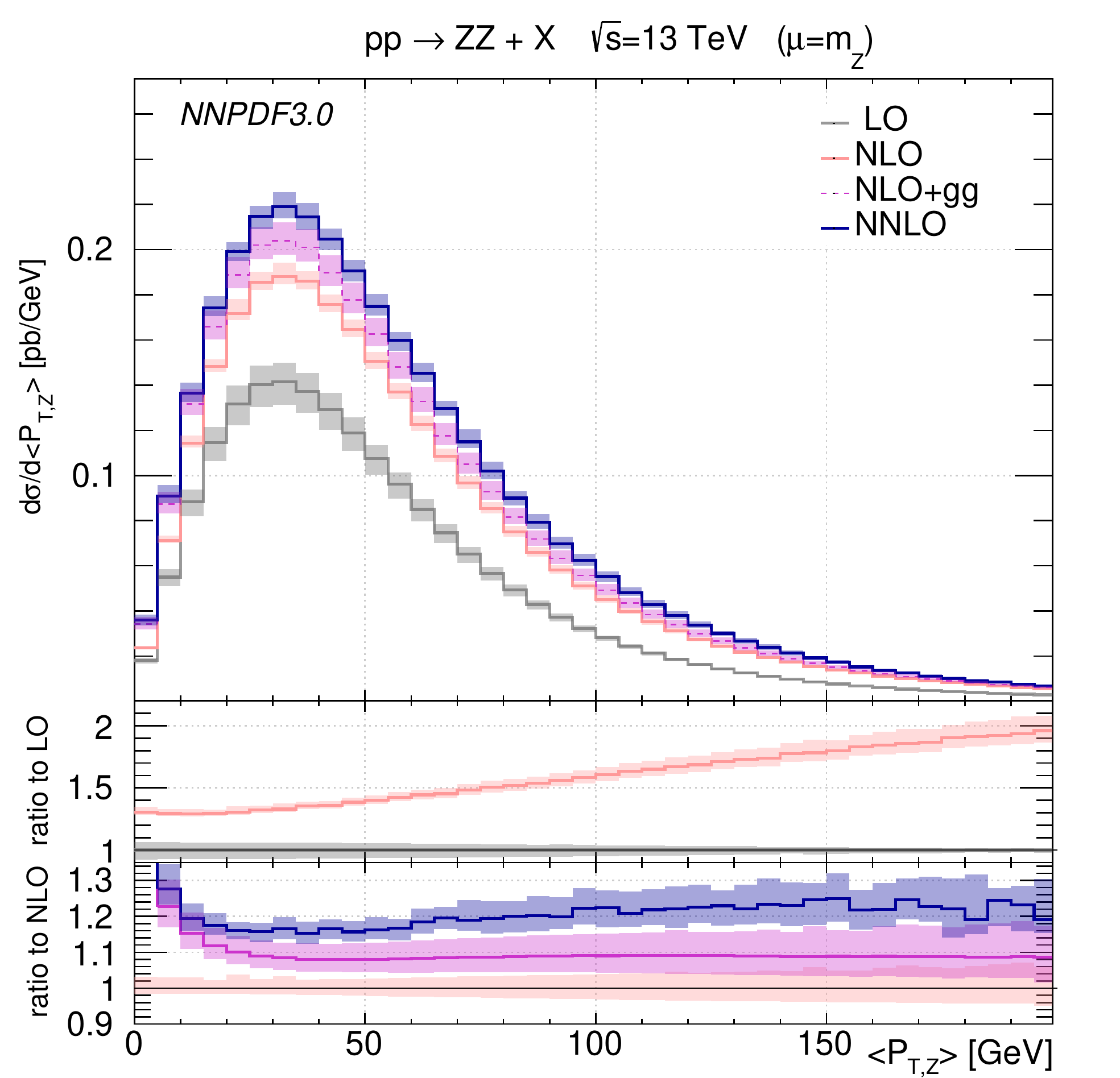}
  \caption{}
  \label{fig:diffxssub2}
\end{subfigure}
\caption{(a) $ZZ$ invariant mass distribution and (b) averaged transverse momentum distribution $\pTZ$ of the $Z$-bosons computed at LO, NLO and NNLO. In the two sub panels
we show respectively the NLO/LO and NNLO/NLO $K$-factors to visualise the size of the higher order effects. 
The result for the contribution from the  loop-induced $gg\to ZZ$ subset 
of the full NNLO correction is also shown separately. 
Shaded bands represent the theory uncertainty due to the variation of the factorisation and renormalisation scales.}
\label{fig:diffxSec}
\end{figure}

In order to study in more detail the scale uncertainty of the cross section we present in Fig.~\ref{fig:scale} the 
renormalisation and factorisation scale dependence of the $ZZ$ cross section at LO, NLO and NNLO. The figure shows largely non-overlapping
scale uncertainty bands which demonstrate that for this process, the scale variations are insufficient to estimate
missing higher order terms in the perturbative expansion. This however is not unexpected since $ZZ$ production at the LHC
is an electroweak process which exhibits no renormalisation scale dependence at LO. For this reason we obtain large NLO QCD  corrections
to the cross section which are outside the LO scale band. Moreover, when going from NLO to NNLO, the loop-induced
gluon fusion channel $gg \to ZZ$ opens up, and due to the large gluon flux it represents a numerically significant contribution.
Since this new channel contributes for the first time at NNLO its contribution cannot be captured by the scale variation
of the NLO cross section. 
Therefore, when increasing the perturbative order, 
we can observe a systematic reduction of the factorisation scale dependence of the cross section 
(indicated by the thickness of the scale uncertainty band), while there is no
significant reduction of the renormalisation scale dependence. 
To show that this effect can be attributed to the gluon fusion channel opening up at NNLO, we also show the NNLO result excluding this channel, leading to an improved convergence of the perturbative expansion.

The appearance of new channels that open up at NNLO and their importance in the various kinematic regions can be studied
by considering differential results. 
Due to the observed mild power corrections in this process we chose to fix the value
of the $0$-jettiness slicing parameter to $\tauzerocut=10^{-2}$ GeV for all our histograms. In Fig.~\ref{fig:diffxSec} we present the invariant mass of the $ZZ$ system
and the average transverse momentum distribution $\pTZ$ of any $Z$-boson, defined as $\pTZ=(|p_T^{Z_1}|+|p_T^{Z_2}|)/2$.
We also present results for the loop-induced $gg\to ZZ$ channel. 

In Fig.~\ref{fig:diffxssub1} we show our results for the $ZZ$ invariant mass. In the first and second sub-panels
we show the effect of the NLO and NNLO corrections, respectively. We observe in the first sub-panel  large NLO QCD corrections 
which vary between 40\% at low $m_{ZZ}$ and  60\% at high $m_{ZZ}$, and change both the shape
and normalisation of the predicted cross section with respect to the LO result. Going to NNLO we observe an approximately flat increase of the cross section of about 18\% with respect to the NLO result, where
approximately 60\% of this effect comes from the loop-induced $gg\to ZZ$ channel, which is outside the scale uncertainty band of the NLO prediction. Similarly, 
in the transverse momentum distribution (Fig.~\ref{fig:diffxssub2}),  we observe large NLO corrections of approximately 30\% at low $\pTZ$, which
can reach almost 100\% at high $\pTZ$. The shape of the NNLO corrections in the second sub-panel  largely follows the contribution of the loop-induced $gg\to ZZ$ channel
at low $\pTZ$, and we observe a 30\% effect at low $\pTZ$ which decreases to about 18\% at high $\pTZ$. For both distributions we observe good convergence of the
perturbative expansion, however the scale uncertainty bands do not overlap between the orders in the perturbative expansion that we have computed. These results show that
the inclusion of NNLO effects in $ZZ$ production at the LHC is essential to obtain a reliable theoretical description of this process.

\section{Conclusions}
In this work we have calculated the NNLO QCD corrections to on-shell $Z$-boson pair production.
Our calculation of the real emission contributions uses $N$-jettiness to isolate the infrared divergent contributions. We discussed our setup in some detail, also showing a comparison 
between results based on antenna subtraction and results based on $N$-jettiness for the NLO corrections. 

After the inclusion of the NNLO correction in the theory prediction we found good agreement with the results of the recent ATLAS and CMS measurements.
Due to the fact that the numerically sizeable loop-induced $gg\to ZZ$ contribution appears for the first time at NNLO, the scale uncertainties in the NNLO  
prediction are not reduced with respect to NLO. The NNLO corrections increase the NLO result by about 18\%, where
almost 60\% of this increase stems from the loop-induced $gg\to ZZ$ channel.
In view of the numerical importance of this channel, 
it is desirable to add the two-loop diagrams, including massive top quark loops, 
to this channel, which will be left for a subsequent publication.

\section*{Acknowledgements}
We thank Frank Tackmann, Simone Alioli and Max Stahlhofen for useful discussions on $N$-jettiness subtraction.
This research was
supported in part by the Research Executive Agency (REA) of the
European Union under the Grant Agreement PITN-GA2012316704 (HiggsTools).
We also acknowledge support by the Munich Institute for Astro- and Particle Physics (MIAPP)
of the DFG cluster of excellence ``Origin and Structure of the Universe''.
We are also grateful for support and resources
provided by the Max Planck Computing and Data Facility (MPCDF).

\bibliographystyle{JHEP}
\bibliography{refsZZ}

\end{document}